 \documentclass[smallabstract,smallcaptions]{dccpaper}

\usepackage{epsfig}
\usepackage{amsmath}
\usepackage{amssymb}
\usepackage{color}
\usepackage{url}
\usepackage{hyperref}

\newlength{\figurewidth}
\newlength{\smallfigurewidth}

\DeclareMathOperator*{\argmin}{argmin}
\usepackage{physics}
\usepackage{algorithm}
\usepackage{algpseudocode}

\begin{document}

\title
{\large
\textbf{Variable-Rate Learned Image Compression with Multi-Objective Optimization and Quantization-Reconstruction Offsets}
}

\author{%
Fatih Kamisli$^{\ast}$, Fabien Racap\'e$^{\ast}$, and Hyomin Choi$^{\ast}$\\[0.5em]
{\small\begin{minipage}{\linewidth}\begin{center}
\begin{tabular}{c}
$^{\ast}$InterDigital - AI Lab  \\
\url{{fatih.kamisli, fabien.racape, hyomin.choi}@interdigital.com}
\end{tabular}
\end{center}\end{minipage}}
}

\maketitle
\thispagestyle{empty}

\begin{abstract}
Achieving successful variable bitrate compression with computationally simple algorithms from a single end-to-end learned image or video compression model remains a challenge. Many approaches have been proposed, including conditional auto-encoders, channel-adaptive gains for the latent tensor or uniformly quantizing all elements of the latent tensor. This paper follows the traditional approach to vary a single  quantization step size to perform uniform quantization of all latent tensor elements. However, three modifications are proposed to improve the variable rate compression performance. First, multi objective optimization is used for (post) training. Second, a quantization-reconstruction offset is introduced into the quantization operation. Third, variable rate quantization is used also for the hyper latent. All these modifications can be made on a pre-trained single-rate compression model by performing post training. The algorithms are implemented into three well-known image compression models and the achieved variable rate compression results indicate negligible or minimal compression performance loss compared to training multiple models. (Codes will be shared at \href{https://github.com/InterDigitalInc/CompressAI}{https://github.com/InterDigitalInc/CompressAI})
\end{abstract}

% -----------------------------------------------------------------------------
\section{Introduction} \label{sec:intro}
End-to-end learned image and video compression has attracted great attention with recent works achieving remarkable success \cite{he2022elic, li2023neural}, rivaling or exceeding state-of-the-art standard codecs such as VVC or AV1 \cite{chen2018overview} in compression performance. One of the challenges hindering deployment of learned compression systems in real applications is the achievement of variable bitrate compression with computationally simple algorithms. While varying the compression bitrate can be performed using simple and well-established methods in traditional image and video compression systems, this is not the case for end-to-end learned image or video compression systems. Indeed, in the initially proposed learned compression systems \cite{balle2017end, balle2018variational, minnen2018joint}, this problem was not addressed and multiple sets of weights for the neural networks (NN) were trained, each targeting a particular trade-off between rate and distortion.%%, i.e. compression rate. % However, storing and using multiple sets of NN weights at the encoder and the decoder is not a feasible solution to varying the bitrate in practical learned compression systems.

To address this issue, many approaches have been proposed %%\cite{choi2019variable, yang2020variable, lin2021deeply, song2021variable, yin2022universal, cui2021asymmetric, tong2023qvrf}. 
\cite{choi2019variable, lin2021deeply, cui2021asymmetric, tong2023qvrf}. These approaches typically use a single set of weights for the NNs in the compression system (i.e., a single NN model) and introduce additional parameters and/or algorithms to vary the compression bitrate. One group of algorithms are based on conditional auto-encoders, where the analysis and synthesis functions of the auto-encoder also take a bitrate alteration parameter %%, such as the $\lambda$ used to form the rate-distortion training cost, 
as input and process it with additional neural networks to alter the output of one or more convolution layers in the analysis and synthesis networks \cite{choi2019variable, lin2021deeply}. %%\cite{choi2019variable, yang2020variable, lin2021deeply, song2021variable}. %%However, such approaches incur complicated designs and introduce additional neural networks and computational complexity. 
Another group of algorithms modifies only the latent tensor prior to (and after) the rounding operation by multiplying or dividing the latent tensor with gain factors or quantization step sizes \cite{cui2021asymmetric, tong2023qvrf}. %% In \cite{cui2021asymmetric}, each latent tensor channel is multiplied by its own gain factor, before and after the rounding operation, and the two factors are not necessarily the inverses of each other. In \cite{tong2023qvrf}, each latent tensor channel is divided by the same (i.e. independent of channel) quantization step-size before rounding, and multiplied by it after rounding, which is quite similar to the quantization process used in traditional image and video processing.

This paper follows the traditional approach, used also by Tong et al. \cite{tong2023qvrf}, and varies a single quantization step size to perform uniform quantization of the latent tensor to achieve variable-rate learned compression. However, three modifications are proposed to improve the variable-rate compression performance. First, multi objective optimization \cite{sener2018multi, desideri2012multiple} is used for (post) training instead of the standard single-objective training procedure used by many variable rate compression algorithms. Second, a quantization-reconstruction offset is introduced into the quantization operation. Third, quantization with variable size quantization step is used also for the hyper latent tensor, which is used in many VAE-based compression approaches. In addition, all of these modifications can be made on a pre-trained single-rate NN model by performing post training. The algorithms are implemented into three well-known image compression models \cite{balle2018variational, minnen2018joint} and the achieved variable rate compression results indicate negligible or minimal compression performance loss, depending on bitrate and model architecture, compared to the results from training multiple NN models.

The remainder of the paper is organized as follows. Section \ref{sec:relwork} provides an overview of related work. Section \ref{sec:prop} presents and discusses the three proposed algorithms. Section \ref{sec:expres} presents experimental results and Section \ref{sec:conc} concludes the paper.

% -----------------------------------------------------------------------------
\section{Related Work} \label{sec:relwork}
%%This section reviews major variable bitrate compression algorithms from the literature in Section \ref{ssec:vbr} and accompanying training algorithms in Section \ref{ssec:vbrtr}.

\subsection{Algorithms for variable bitrate compression with a single set of NN weights} \label{ssec:vbr}
% Varying the compression rate of the encoded content is a fundamental requirement of any lossy image or video compression system. For example, the encoder may want to increase the compression ratio to reduce the bitrate of the compressed bitstream to accommodate the transmission rate requirements of a communication medium. Alternatively, the encoder may want to decrease the compression ratio to allow the decoder to reconstruct an image or video content with lower distortion. The desired compression bitrate may be set prior to starting the compression or during compression. For example, in many video compression applications, each frame of the content may need to be compressed with a slightly different compression ratio to prevent large fluctuations of the bitrate or distortion of successive frames. 

Varying the compression rate of the encoded content is a fundamental requirement of any lossy image or video compression system. 
Traditional image or video compression systems change a single scalar parameter, the quantization step-size, used in quantizing transform coefficients in the compression system. Most of the parameters and algorithms within the compression system do not need to be changed. The advantages of this method are that it provides a simple and computationally efficient algorithm %%with an easily predictable outcome 
to vary the compression ratio, while also providing an optimal or close-to-optimal trade-off between the bitrate and distortion as the compression ratio is changed. %%The latter property is mainly due to the used transforms being so-called orthonormal transforms having special properties.

%%While varying the compression ratio or bitrate can be performed using simple and well-established methods for traditional image and video compression systems, this is not the case for end-to-end learned image and video compression systems. Indeed, in the initially proposed end-to-end learned compression systems \cite{balle2017end, balle2018variational, minnen2018joint}, this issue was not addressed and multiple sets of weights for the neural networks (NN) were trained, each targeted for a particular trade-off between rate and distortion, i.e. a compression rate.

Many approaches have been proposed to perform variable bitrate image or video compression with a single set of NN weights (i.e., a single NN model). %%\cite{choi2019variable, yang2020variable, lin2021deeply, song2021variable, yin2022universal, cui2021asymmetric, tong2023qvrf}. 
One group of approaches are based on conditional auto-encoders \cite{choi2019variable, lin2021deeply} and another is based on processing only the latent tensor prior to and after the rounding operation \cite{cui2021asymmetric, tong2023qvrf}. %%One specific conditional auto-encoder algorithm \cite{choi2019variable} and two latent tensor processing algorithms \cite{cui2021asymmetric, tong2023qvrf} are discussed in some detail below.

%%In \cite{choi2019variable}, the authors propose a variable bitrate image compression framework based on a conditional auto-encoder, which includes conditional convolution layers (conditioned on the trade-off or Lagrange parameter $\lambda$ in the rate-distortion training cost $R+\lambda D$ and the quantization step-size) instead of the standard convolution layers. They have thus two parameters to vary the bitrate. Coarse rate adaptation to a target is performed by changing $\lambda$, while the rate can be further fine-tuned by adjusting the bin size used in quantization of the latent. Overall, this is a complicated design/algorithm since the conditional convolution layers are more computationally complex than standard convolution layers.

\begin{figure}[t]
\centering
\includegraphics[clip,trim=0 10 0 1,width=0.61\linewidth]{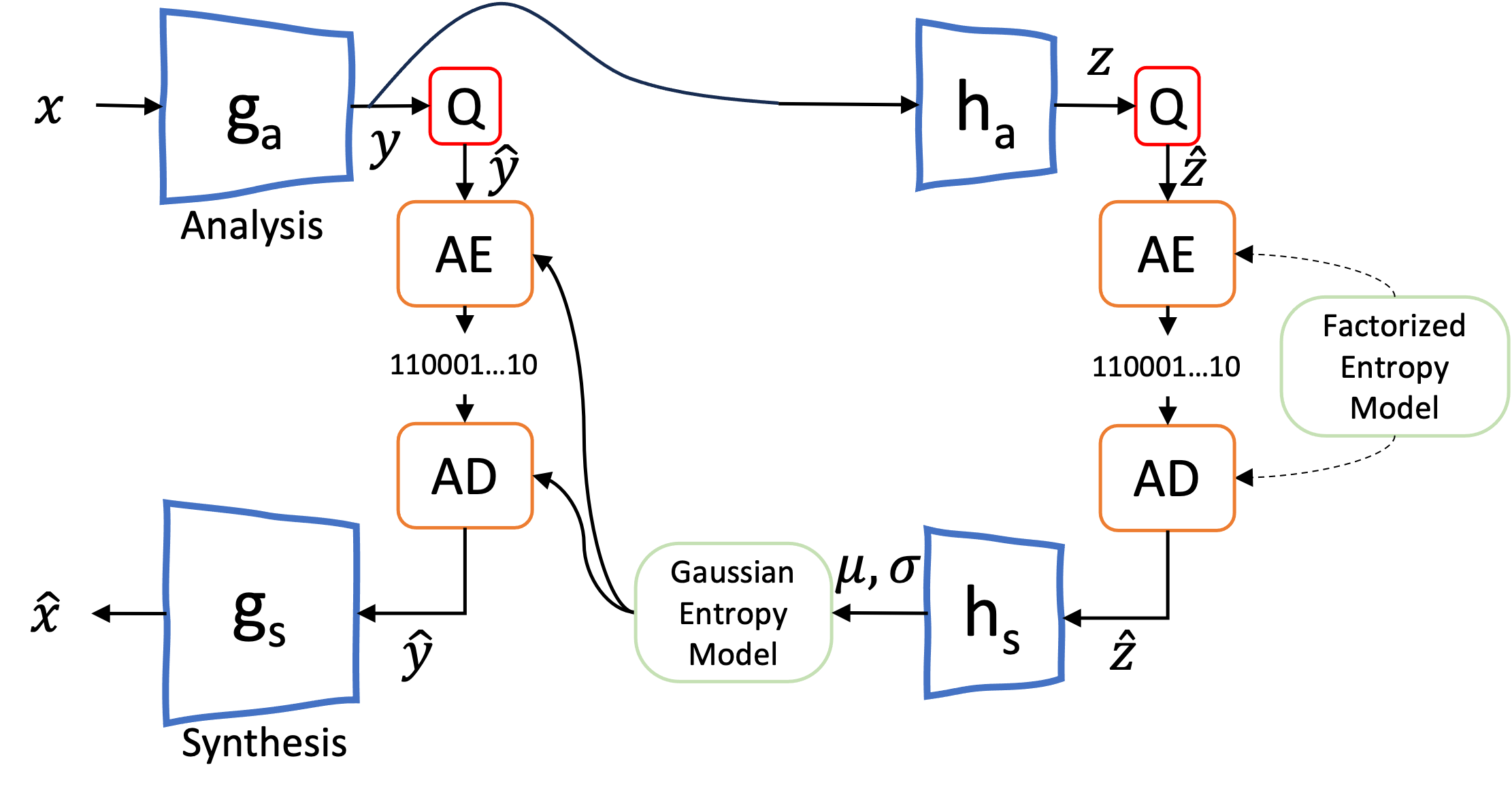}
\vspace{-0.1in}
\caption{\label{fig:mshp}
Mean\&Scale Hyperprior based image compression model \cite{minnen2018joint}.}
\end{figure}

In \cite{cui2021asymmetric}, the authors propose a variable bitrate image compression framework based on so-called gain units. Gain units are simply multiplication factors per latent tensor channel. In other words, each latent tensor channel is multiplied by its own two factors, before and after the rounding operation, and the two factors are not necessarily the inverses of each other:
\begin{align}
  q_{c,i,j} &= round(y_{c,i,j}\cdot G_c) \label{eq:cui_1} \\
  \hat{y}_{c,i,j} &= q_{c,i,j} \cdot G_c^{'}  \label{eq:cui_2}
\end{align}
In the first equation above, $y_{c,i,j}$ represents the latent variable (in channel $c$ at horizontal position $i$ and vertical position $j$) output by the analysis function $g_a$ of the learned compression system (see Figure \ref{fig:mshp}). $G_c$ is the gain or multiplication factor for latent channel $c$ applied before rounding, and $round(.)$ is the operation of rounding the value to the nearest integer. $q_{c,i,j}$ is the resulting quantization index. In the second equation, the reconstruction of the quantized value $\hat{y}_{c,i,j}$ of the latent variable %%from the quantization index $q_{c,i,j}$ 
is performed by multiplying again with a gain $G_c^{'}$. 

To adjust the bitrate coarsely, a predefined set of six $G_c$  and $G_c^{'}$ factors are learned for each latent channel and to adjust the rate more finely, two of these predefined factors are chosen and interpolated to obtain the desired gain units. Overall, this is a simpler design/algorithm than the ones based on conditional auto-encoders. The presented variable rate compression results of this method \cite{cui2021asymmetric} indicate that, compared to training multiple NN models, similar or better compression performance is achieved at low bitrates while inferior compression performance is achieved at higher bitrates.

In \cite{tong2023qvrf}, Tong et al. propose a variable rate image compression framework based on varying a single parameter, the quantization step-size $\Delta$, that is used to quantize all latent tensor variables as follows\footnote{Notice that while the authors use a so-called quantization regulator parameter $a$, their processing is essentially identical to those in Equations \ref{eq:qvrf_1}$\&$\ref{eq:qvrf_2} with $a=\frac{1}{\Delta}$ \cite{tong2023qvrf}}:
\begin{align}
  q_{c,i,j} &= round(\frac{y_{c,i,j}}{\Delta}) \label{eq:qvrf_1} \\
  \hat{y}_{c,i,j} &= q_{c,i,j} \cdot \Delta \label{eq:qvrf_2}
\end{align}

These equations are a special case of the above Equations \ref{eq:cui_1}$\&$\ref{eq:cui_2}. If $G_c=\frac{1}{\Delta}$, and $G_c{'}=\Delta$, the same equations are obtained. Note that $\Delta$ is now independent of the channel index $c$. %%To adjust the bitrate, this quantization step-size $\Delta$ is varied. 
This method simply comes down to the uniform quantization process. It is used in many traditional signal processing algorithms including traditional image and video compression. While this method provides an even simpler design/algorithm than the one in \cite{cui2021asymmetric}, its compression performance is also inferior at higher bitrates (up to 1dB), compared to results from training multiple NN models \cite{tong2023qvrf}.

\subsection{Training of variable bitrate compression NN models} \label{ssec:vbrtr}
First, the training of single bitrate NN models is reviewed. Then the typical approach in the literature to train the variable bitrate compression NN models is discussed.

NN models that target a single bitrate are trained using a loss function $L$, which consists of a Lagrangian form obtained from the bitrate $R$ and the distortion $D$ ($L=R+\lambda D$). Both $R$ and $D$ are functions of the NN parameters $\theta$. The training is performed with stochastic gradient descent to minimize the loss function with respect to the parameters $\theta$ over a training set \cite{balle2017end}:
\begin{align}
    \arg \min_\theta R(\theta) + \lambda D(\theta) \label{eq:rdc}
\end{align}
% Here, $R(\theta)$ and $D(\theta)$ are the bitrate and the distortion of the decoded content, and both are functions of the NN parameters $\theta$. 
$\lambda$ is a factor that determines the trade-off between rate and distortion and thus also the target bitrate \cite{balle2017end, balle2018variational}. For different target bitrates, a new model (i.e., new set of NN weights) is trained with an appropriate $\lambda$ in the loss function. %%Thus, if one uses NN models that target a single bitrate, then, in a practical compression application where the target bitrate needs to be changed, multiple NN models (i.e., sets of weights for the NN) need to be stored at the encoder and decoder, which is not practical.

Variable bitrate compression systems with a single NN model, such as those discussed in Section \ref{ssec:vbr}, employ a slightly different training strategy \cite{choi2019variable, cui2021asymmetric, tong2023qvrf}. They minimize a loss function that consists of the sum of the loss functions with $N$ different $\lambda$ values, which are chosen for a range of appropriate target bitrates:
\begin{align}
    & \argmin_{\theta, \phi_1,...,\phi_N} \sum_{i=1}^N R(\theta, \phi_i) + \lambda_i D(\theta, \phi_i)  \label{eq:vrrdc}
    % \text{where} \quad & L_i(\theta) =  , \quad i=1,...,N  \label{eq:vrrdc}
\end{align}
When calculating the loss for the $i^{th}$ target bitrate ($L_i(\theta, \phi_i)=R(\theta, \phi_i) + \lambda_i D(\theta, \phi_i)$), the corresponding bitrate variation parameters $\phi_i$ are used, i.e., corresponding gains $G_c^{(i)}$ and $G_c{'}^{(i)}$ in \cite{cui2021asymmetric} or a corresponding $\Delta^{(i)}$ in \cite{tong2023qvrf}. %%Note that since the cost in Equation \ref{eq:vrrdc} requires $N$ forward passes through the NN, the training time increases and thus an alternative/simpler training algorithm is preferred in \cite{tong2023qvrf}, where for each batch of the training, one out of $N$ target bitrates is chosen at random and a single forward and backward pass over the NN are performed using only the loss for this target bitrate, instead of the sum of losses.

% -----------------------------------------------------------------------------
\section{Proposed Methods} \label{sec:prop}
This paper follows the approach of Tong et al. \cite{tong2023qvrf} and varies a single quantization step-size $\Delta$ to perform the uniform quantization of the latent tensor to achieve variable-rate learned image compression. However, three modifications are proposed to improve the variable rate compression performance. First, multi-objective optimization is used for (post) training instead of the standard single-objective optimization algorithm. Second, a quantization-reconstruction offset is introduced into the quantization operation. Third, quantization with variable quantization step-size is used also for the hyper latent tensor. Each of these algorithms is discussed below.

\subsection{Multi-objective optimization (MOO)} \label{ssec:moo} 
% Consider two different parameter sets $\Theta=[\theta,\phi_1,...,\phi_N]$ and $\bar{\Theta}=[\bar{\theta},\bar{\phi}_1,...,\bar{\phi}_N]$ for a variable rate compression system. These parameter sets are such that $L_i(\theta, \phi_i)<L_i(\bar{\theta}, \bar{\phi}_i)$ and $L_j (\theta, \phi_j)>L_j (\bar{\theta}, \bar{\phi}_j)$. In other words, for the $i^{th}$ target bitrate, the first parameter set gives a better compression result while for the $j^{th}$ target bitrate, the other parameter set gives a better result. To determine which parameter set is a better choice for the variable compression rate system requires some sort of importance weighting of the losses, however, the optimal weighting is an open question. The formulation of the loss function in Equation \ref{eq:vrrdc} that consists of sum of losses $L_i(\theta,\phi_i)$ places equal importance on all losses, which may not be a good choice.

Multi-objective optimization (MOO) proposes an alternative formulation to the variable bitrate end-to-end learned image or video compression problem. In MOO, a collection of possibly conflicting objectives, in our case the loss functions $L_i(\theta, \phi_i)$ for different target bitrates, are optimized together by using a vector-valued loss:
\begin{align}
    & \argmin_{\theta, \phi_1,...,\phi_N}~ [L_1(\theta_{},\phi_1),~ L_2(\theta_{},\phi_2),~ ... ,~ L_N(\theta_{},\phi_N)]  \label{eq:moo}
\end{align}
Here, all parameters of the variable rate compression system are represented in two groups. $\theta_{}$ represents the parameters of the NN that are shared/common when performing compression for any bitrate, which is typically a great majority of the NN parameters. $\phi_i$, $i=1,...,N$ represent the parameters that are specific to a target bitrate, i.e., used only during compression for the $i^{th}$ target bitrate. For example, for the image compression system in Figure \ref{fig:mshp}, $\theta_{}$ would include all parameters/weights of $g_a, g_s, h_a, h_s$ and the factorized entropy model (as in a single target bitrate model), while $\phi_i$ would include parameters such as the gains $G_c$, $G_c{'}^{(i)}$ in \cite{cui2021asymmetric} or the $\Delta$ in \cite{tong2023qvrf}.% or the $\Delta$ and quantization-reconstruction offsets in our framework.

A solution to such an MOO problem is proposed to be a set of NN parameters $\Theta=[\theta_{},\phi_1,...,\phi_N]$ which achieves so-called Pareto optimality, defined as follows \cite{sener2018multi, desideri2012multiple}:
\begin{itemize}
    \item A solution $\Theta=[\theta_{},\phi_1,...,\phi_N]$ dominates a solution $\bar{\Theta}=[\bar{\theta}_{},\bar{\phi}_1,...,\bar{\phi}_N]$ if $L_i(\theta_{}, \phi_i)\leq L_i(\bar{\theta}_{}, \bar{\phi}_i)$ for all $i=1,...,N$.
    \item A solution $\Theta^*$ is Pareto optimal if there exists no solution that dominates $\Theta^*$.
\end{itemize}

The MOO problem can be solved using different optimization algorithms, including gradient based algorithms. In \cite{desideri2012multiple, sener2018multi}, the Multiple gradient descent algorithm (MGDA) is proposed for MOO. The following problem is utilized in MGDA:
% \begin{align}
%     \min_{\alpha_1,\alpha_2,...,\alpha_N}~ & \norm{ \sum_{i=1}^N\alpha_i \nabla_{\theta_{}}L_i(\theta_{},\phi_i) }_2^2 \label{eq:norm}  \\ 
%     \text{s.t.}~& \alpha_1+\alpha_2+...+\alpha_N=1 \nonumber \\ 
%     & \alpha_i\geq 0 ~~\text{for}~ i=1,...,N \nonumber
% \end{align}
\begin{align}
    \min_{\alpha_1,\alpha_2,...,\alpha_N}~ & \norm{ \sum_{i=1}^N\alpha_i \nabla_{\theta_{}}L_i(\theta_{},\phi_i) }_2^2 \label{eq:norm}  \\ 
    \text{s.t.}~~ \sum&_{i=1}^N \alpha_i=1 ~~\text{and}~~ \alpha_i\geq 0 ~\forall i  \nonumber 
\end{align}
This problem seeks the minimum norm solution in the convex hull of the gradient vectors $\nabla_{\theta_{}}L_i(\theta_{},\phi_i)$, $i=1,...,N$. It is claimed that either the solution is 0 and the associated $\Theta$ is a Pareto stationary solution (i.e., a locally Pareto optimal) or the resulting solution $\sum_{i=1}^N\alpha_i \nabla_{\theta_{}}L_i(\theta_{},\phi_i)$ gives a descent direction that improves all losses $L_i$, $i=1,...,N$. Based on this claim, Algorithm \ref{alg:cap} below, which also includes regular gradient descent on non-shared parameters $\phi_i$, is proposed for the MOO problem. (Notice that we will skip how the solution to the optimization problem in Equation \ref{eq:norm} is obtained; one approach based on the Frank-Wolfe method is given in \cite{sener2018multi})
\begin{algorithm}
\caption{MOO for variable bitrate compression \cite{sener2018multi}}\label{alg:cap}
\begin{algorithmic}[1]
\For{$i$ to $N$}
    \State $\phi_i = \phi_i - \eta \nabla_{\phi_i}L_i(\theta_{},\phi_i) $      \qquad \Comment{Gradient descent on bitrate-specific params}
\EndFor
\State $\alpha_1,\alpha_2,...,\alpha_N$=\text{MinNormSolver}$(\nabla_{\theta_{}}L_i(\theta_{},\phi_1),...,\nabla_{\theta_{}}L_N(\theta_{},\phi_N))$
\State $\theta_{} = \theta_{} - \eta \sum_{i=1}^N \alpha_i \nabla_{\theta_{}}L_i(\theta_{},\phi_i)$   \qquad \Comment{Gradient descent on shared params}
\end{algorithmic}
\end{algorithm}

To compare the MOO algorithm with the conventional training algorithm, consider line 5 in Algorithm \ref{alg:cap} and Equation \ref{eq:vrrdc}. Equation $\ref{eq:vrrdc}$ puts equal weight on all losses $L_i$ (and thus their gradients in the parameter update) while line 5 updates the parameters by putting adaptively calculated weights $\alpha_i$ on the gradients of $L_i$.

\subsection{Quantization-Reconstruction (QR) offsets} \label{ssec:qro}
A quantization-reconstruction (QR) offset $\delta$ is introduced into the quantization process within end-to-end learned image or video compression: 
\begin{align}
  q_{c,i,j} &= round(\frac{y_{c,i,j}}{\Delta}) \label{eq:qro_1} \\
  \hat{y}_{c,i,j} &= 
      \begin{cases} 
        (q_{c,i,j}+\delta_{c,i,j}) \cdot \Delta, \quad & \text{if } q_{c,i,j} \neq 0 \label{eq:qro_2}\\
        (q_{c,i,j}+             0) \cdot \Delta, \quad & \text{if } q_{c,i,j} = 0 
      \end{cases}
\end{align}
These equations are similar to Equations \ref{eq:qvrf_1}\&\ref{eq:qvrf_2} except that an offset $\delta_{c,i,j}$ is introduced into the reconstruction equation for non-zero quantization indices. 

\begin{figure}[htb]
\centering
    \includegraphics[clip,trim=0 0 0 0,width=0.43\linewidth]{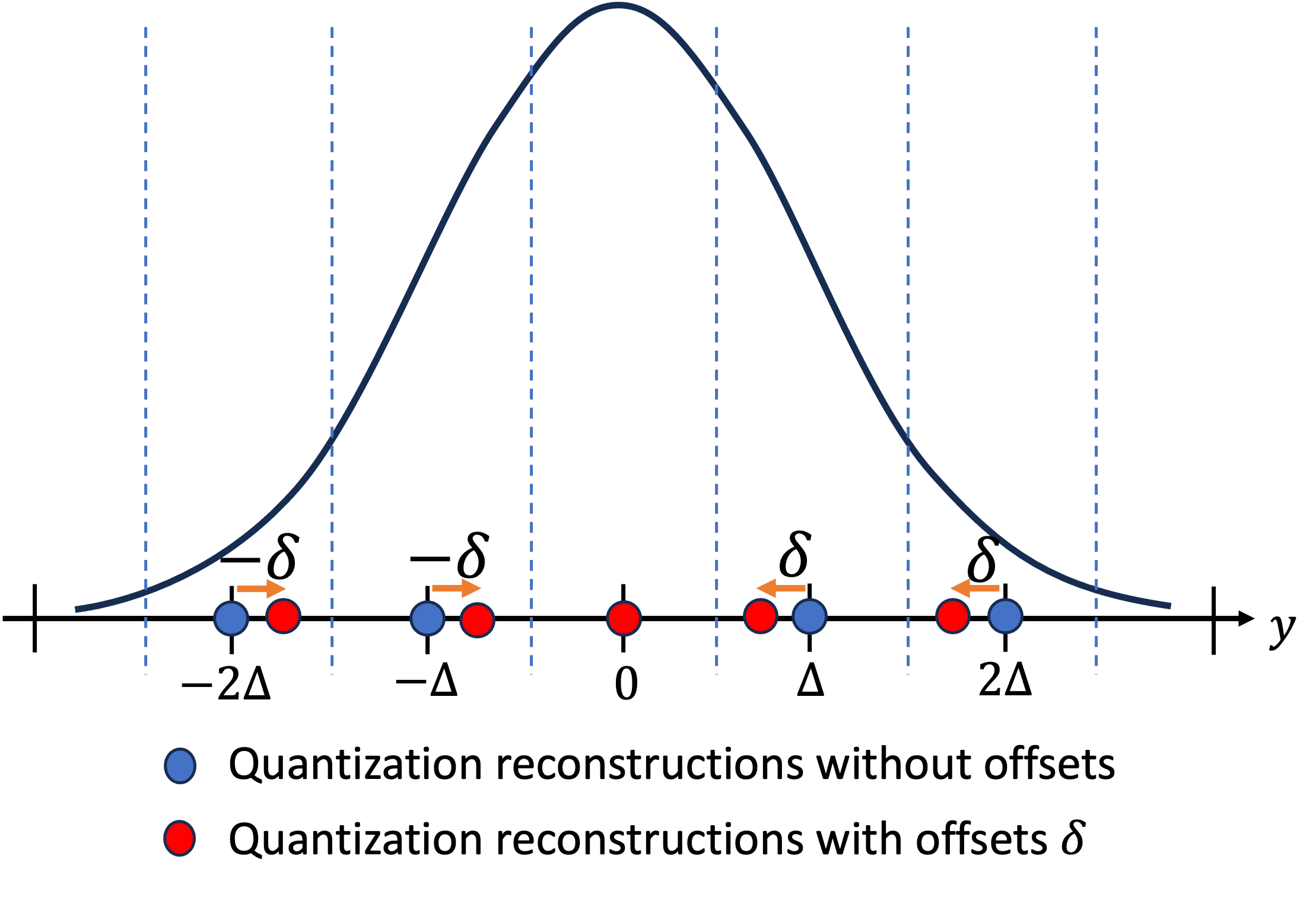} \hspace{0.5in}
    \includegraphics[clip,trim=10 10 10 10,width=0.31\linewidth]{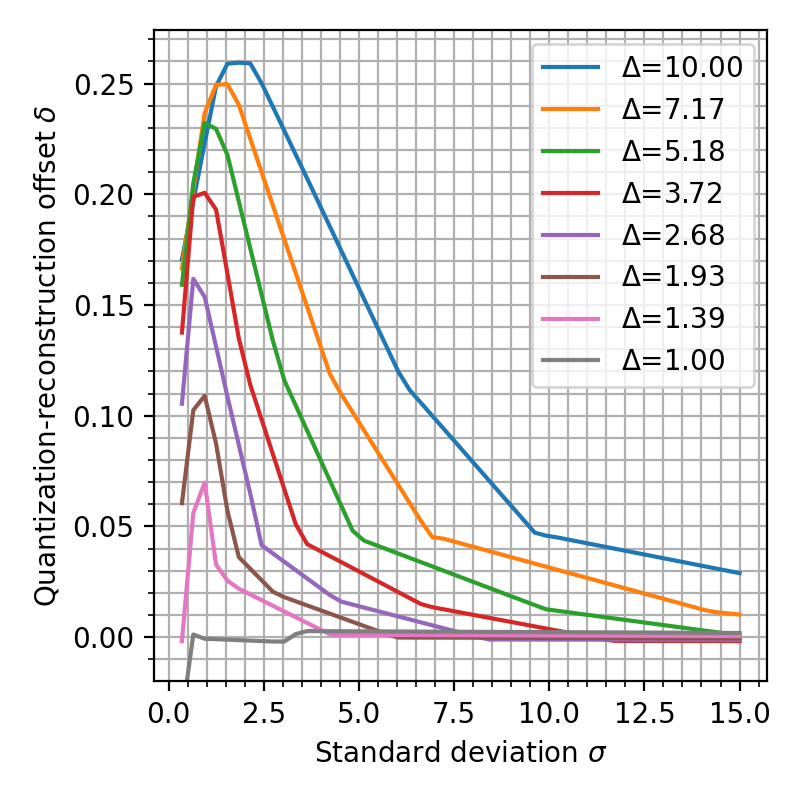} \\
    {\footnotesize (a) Quantization with reconstruction offsets}  \hspace{0.3in}  {\small (b) learned offsets as functions of $\sigma$ and $\Delta$}
\vspace{-0.2in}
\caption{\label{fig:qro} Quantization with learned quantization-reconstruction offsets}
% \caption{\label{fig:qro} (a) Quantization with reconstruction offsets (b) learned offsets as functions of $\sigma$ and $\Delta$}
\end{figure}

The motivation behind using a reconstruction offset in quantization is as follows. For a unimodal distribution such as the Gaussian distribution, which is the assumed (conditional) distribution of the latent $y_{c,i,j}$ in many approaches, using a quantization reconstruction value at the center of each quantization interval is not optimal for the mean-squared-error (MSE) of the reconstructed value (see Figure \ref{fig:qro}-a). Thus, traditional image or video compression systems typically use QR offsets (when quantizing transform coefficients.) However, note that end-to-end learned image compression systems typically do not \cite{balle2017end, balle2018variational, balle2020nonlinear}. The reason behind this is that when such systems are trained for a specific bitrate with a specific $\lambda$ in the loss function, the functionality of the QR offset is implicitly learned by the synthesis function $g_s$ and thus there is no need for an explicit offset term in the quantization reconstruction equation \cite{balle2020nonlinear}. However, in a variable bitrate compression system with a single NN model, the synthesis function cannot learn the functionality of the QR offset for a wide range of quantization step-sizes $\Delta$, and thus using an explicit QR offset becomes now useful and improves variable bitrate compression performance as shown in Section \ref{sec:expres}.

A major question is how to obtain the QR offsets $\delta_{c,i,j}$. Multiple approaches can be envisioned. Considering that a QR offset to minimize the representation MSE is dependent on the skewness of the distribution within a quantization interval, and this skewness can change depending on the quantization step-size $\Delta$ and the distribution variance $\sigma^2$, this paper chooses to compute the QR offsets $\delta_{c,i,j}$ from these parameters. A simple 3-layer fully-connected neural network (with 12 hidden nodes) is found to be sufficient. The input is simply the estimated standard deviation $\sigma_{c,i,j}$ (which is available in most learned compression systems for the entropy model) of the latent $y_{c,i,j}$ and the used quantization step-size $\Delta$. The output is the QR offset $\delta_{c,i,j}$. Figure \ref{fig:qro}-b shows, for a particular compression model, the learned offsets $\delta$ as a function of $\sigma$ and $\Delta$. As expected, as $\sigma$ increases or $\Delta$ decreases, the skewness of the distribution within a quantization interval decreases and the QR offset $\delta$ gets closer to zero.

\subsection{Variable-rate hyper latents}
A large fraction of the total bitrate of a learned image compression model is used for the latent $y$ and the remaining fraction is used for the hyper latent $z$ (see Figure \ref{fig:mshp}). The values of these fractions depend on the architecture of the model and the compression bitrate \cite{balle2018variational}. In variable bitrate compression, as the compression bitrate is changed by varying the quantization of the latent $y$, the bitrate of the hyper latent $z$ should also be adjusted. This is done in \cite{cui2021asymmetric} but not in \cite{tong2023qvrf}. Since this paper builds upon the quantization framework in \cite{tong2023qvrf}, variable rate quantization (similar to Equations \ref{eq:qvrf_1}\&\ref{eq:qvrf_2}) is also performed for the hyper latent $z$ and it is observed that this can improve the variable compression performance slightly. %% as shown in Section \ref{sec:expres}.
The quantization step-sizes $\Delta_{z}$ used for hyper latent $z$ are learned with a simple 3-layer fully-connected NN with one input, the $\Delta$ of the latent $y$, 10 hidden nodes and one output, the $\Delta_z$.

\subsection{Continuously variable bitrate adjustment}
Finally, note that the proposed algorithms allow for continuously variable bitrate adjustment. While the training was performed with eight $\lambda$ and corresponding eight $\Delta$ values, using any quantization step-size $\Delta$ within the range defined by the eight $\Delta$ values (i.e. [1, 10], or even slightly outside that range) is possible during compression and allows for (practically) continuously variable bitrate with good compression performance, as can be seen from the results presented in Section \ref{sec:expres}. Note that, as $\Delta$ is changed to any desired value, the QR offset $\delta$ and the quantization step-size $\Delta_z$ for the hyper latent are also adjusted by the NNs that produce them from $\Delta$.

% -----------------------------------------------------------------------------
\section{Experimental Results} \label{sec:expres}
%%This section provides some experimental details, and presents variable bitrate compression results and comparisons with related work. 

\subsection{Experimental Settings} \label{ssec:expset}
The three algorithms related to variable bitrate compression discussed in Section \ref{sec:prop} were implemented for three well-known image compression models: Scale Hyperprior \cite{balle2018variational} (Bmshj2018-hp), Mean-Scale Hyperprior \cite{minnen2018joint} (Mbt2018-mean) and Joint AR and Hierarchical Priors \cite{minnen2018joint} (Mbt2018). Their implementations in CompressAI \cite{begaint2020compressai} were used.  %%Some common training settings are as follows. 
%%All training was performed with the Vimeo-90k training set \cite{xue2019video}. Randomly cropped patches of 256x256 pixels with a batch-size of 16 were used. The Adam optimizer was used along with a learning rate scheduler to reduce the learning rate to one third of its value with a patience of 5 epochs. 
All training was performed with Vimeo-90k \cite{xue2019video} and randomly cropped patches of 256x256 and a batch-size of 16. The Adam optimizer was used along with a learning rate scheduler to reduce the rate to one third of its value with a patience of 5 epochs. 

\subsection{Compression Results and Comparisons} \label{ssec:comp}
\begin{figure}[t]
\centering
%\begin{tabular}{cc}
    \includegraphics[clip,trim=20 9 36 28,width=0.49\linewidth, height=0.36\linewidth]{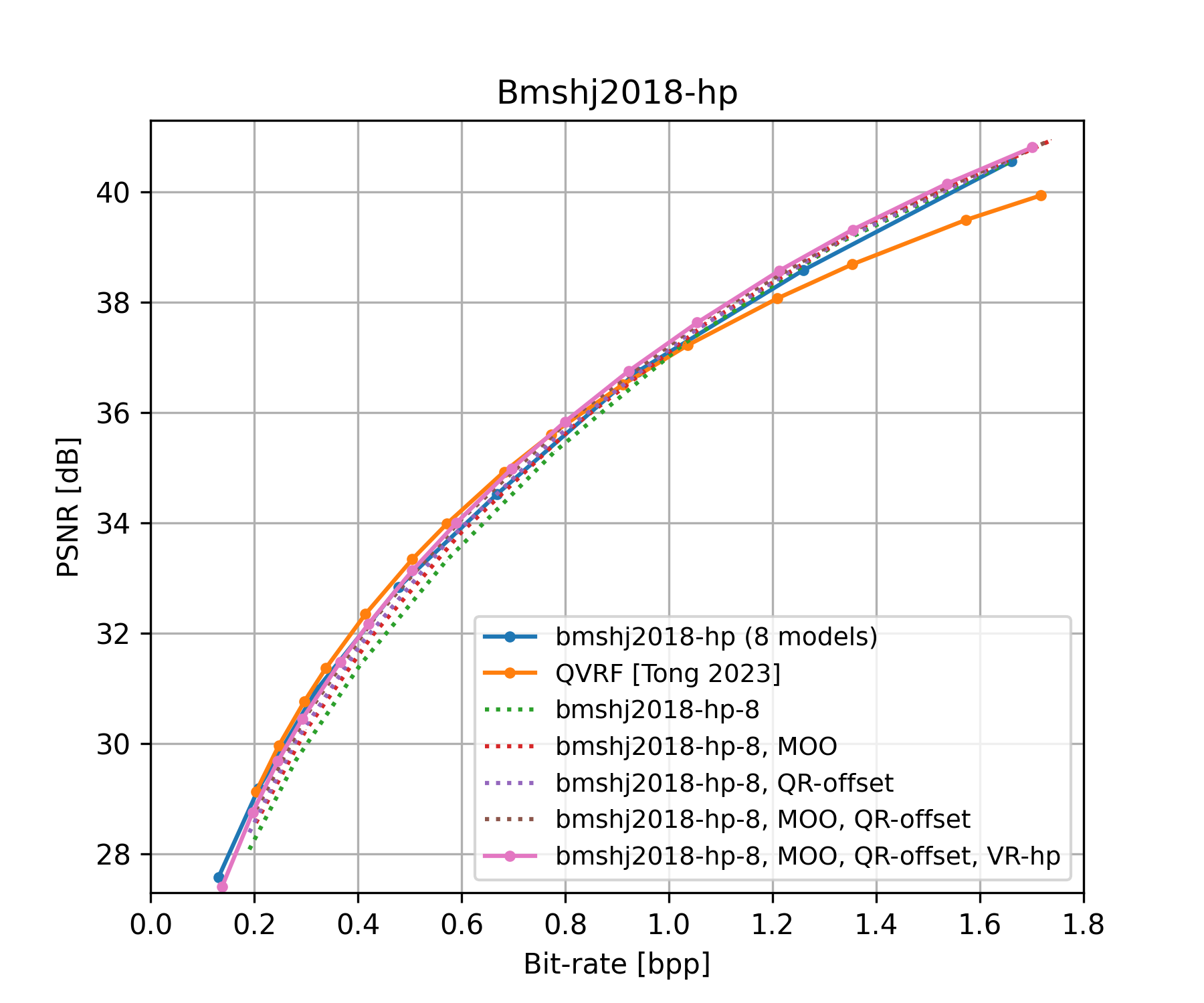} 
    \includegraphics[clip,trim=20 9 36 28,width=0.49\linewidth, height=0.36\linewidth]{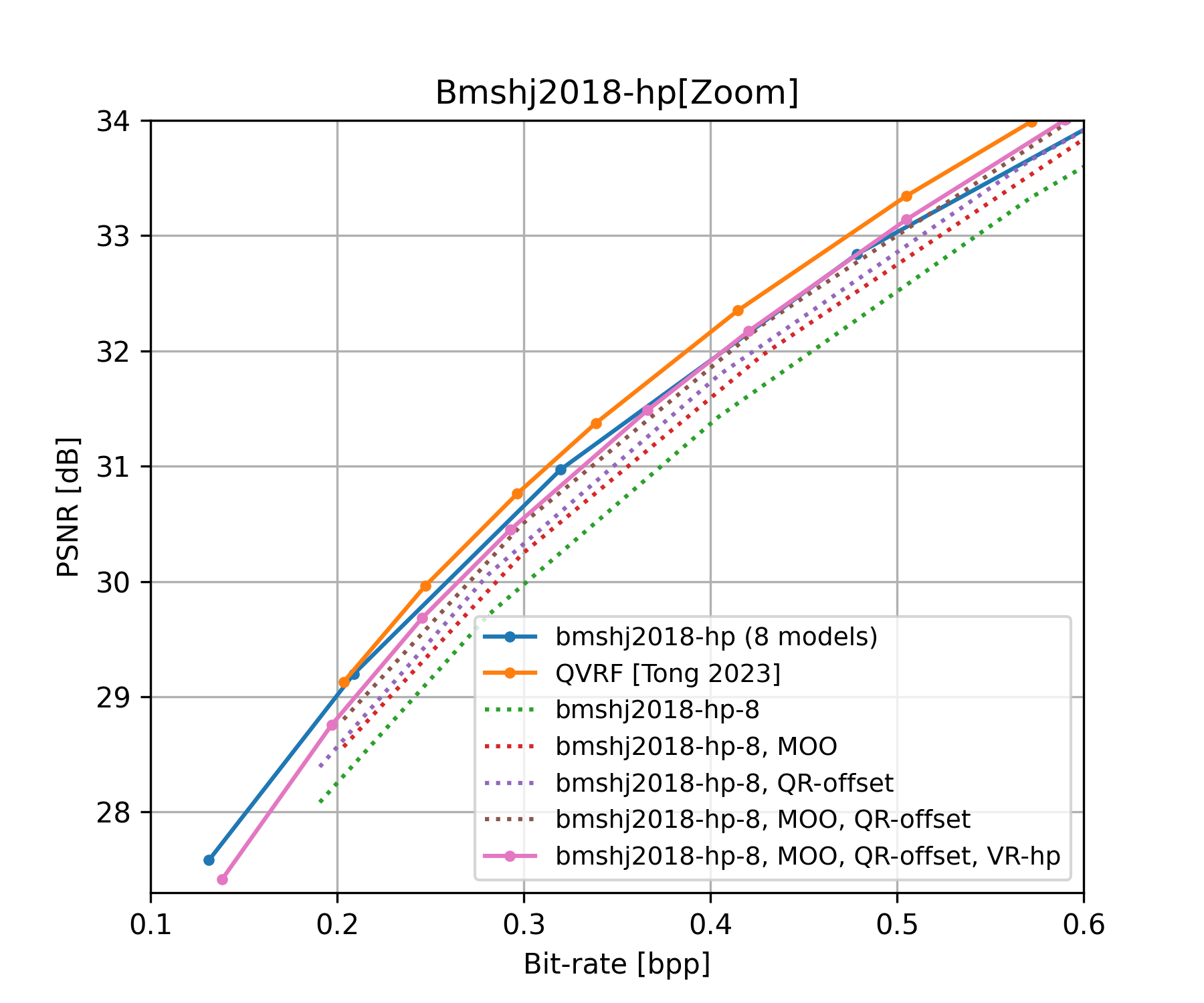}\\ 
    \includegraphics[clip,trim=20 9 36 28,width=0.49\linewidth, height=0.36\linewidth]{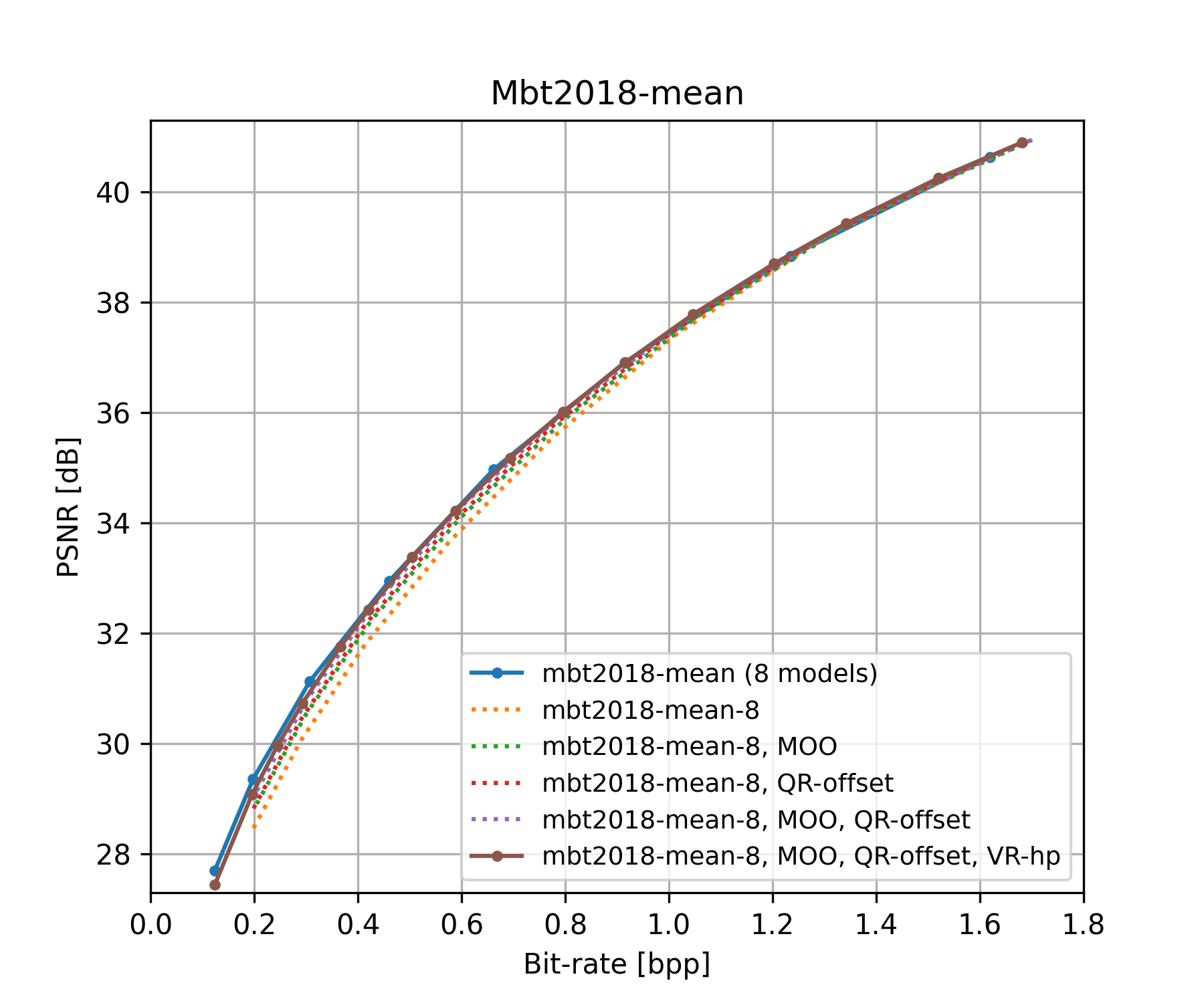} 
    \includegraphics[clip,trim=20 9 36 28,width=0.49\linewidth, height=0.36\linewidth]{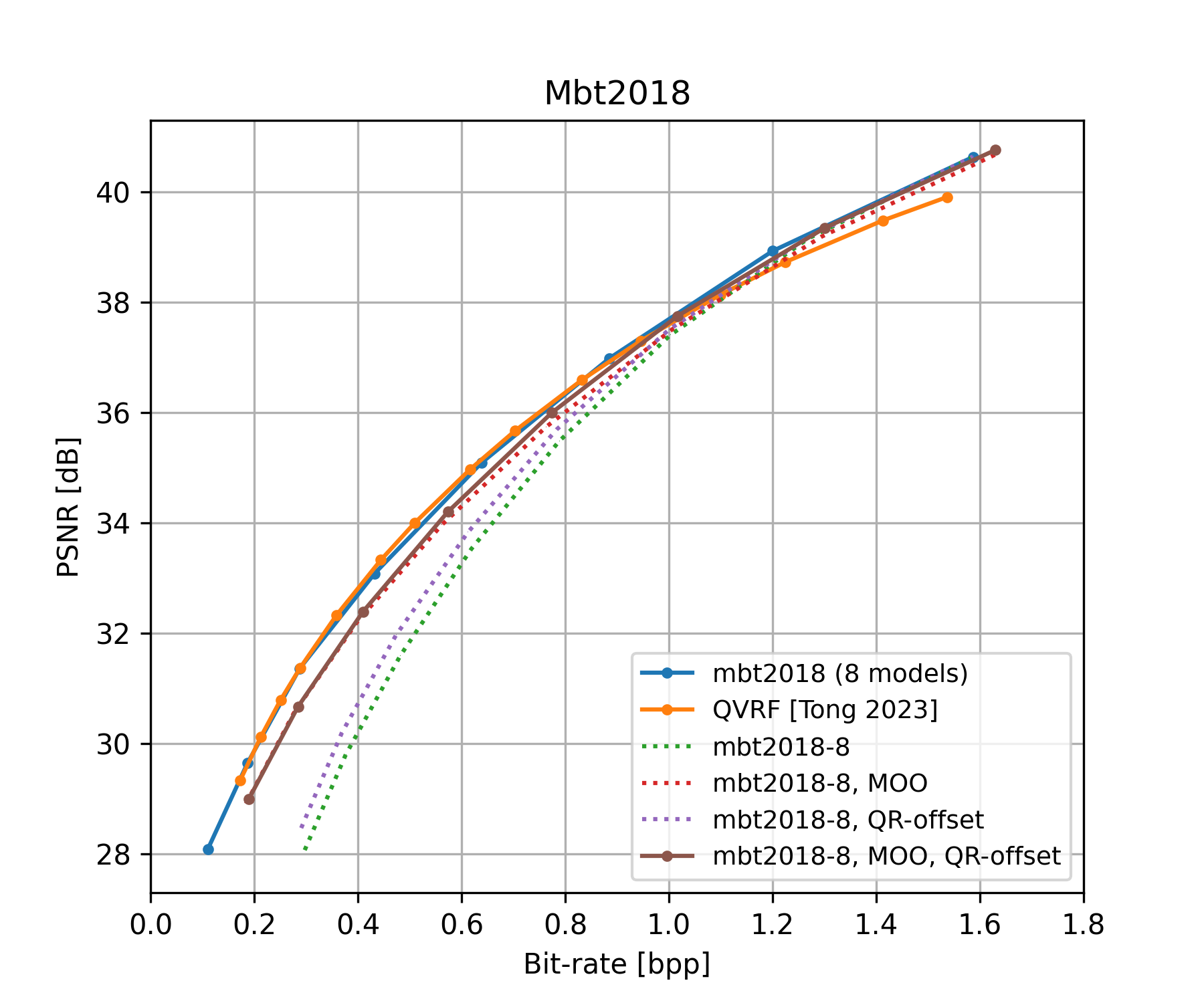} \\
    %{\small (a)} & {\small (b)}
%\end{tabular}
\caption{\label{fig:results} Comparison of compression performance on the Kodak dataset.}% [Starting with top row, left to right.] (a) (b) (c) (d)}
\end{figure}

One question that needs to be asked upfront is as follows: What is the achievable variable bitrate compression performance, if one takes a model that was trained for a single bitrate (in particular with the highest $\lambda$ value) and simply quantizes the latent with larger (than 1) quantization step-sizes $\Delta$. The results are shown in Figure \ref{fig:results} with the dashed curves labeled with the model name and an "-8" at the end (e.g., bmshj2018-hp-8). Compared to the blue curves, which represent the RD performance of 8 models, each trained for a particular bitrate \cite{begaint2020compressai}, there is little PNSR loss at high bitrates but the loss grows reaching about 1dB at the lowest bitrate end for the Bmshj2018-hp and Mbt2018-mean models. This is a surprising result since, first, it requires no special training or bitrate variation algorithms and does not provide catastrophically bad results. Second, the QVRF \cite{tong2023qvrf} result has a similar 1dB loss, but at the high end of the bitrate spectrum. For the Mbt2018 model, however, simply increasing the quantization step-size $\Delta$ provides much poorer results, mainly due to the auto-regressive component in its entropy model \cite{minnen2018joint}.

The training of the image compression models for the three variable bitrate algorithms proposed in Section \ref{sec:prop} is performed in three stages. The first stage does not include the QR offsets or the variable bitrate hyperprior algorithms and only applies the MOO training algorithm with eight $\lambda_i$ values and corresponding quantization step-sizes $\Delta_i=\sqrt{\frac{\lambda_8}{\lambda_i}}$, $i=1,...,8$ \cite{tong2023qvrf}. Note also that for this stage, the image compression model weights are initialized with the weights of the corresponding model from CompressAI that were trained with the highest $\lambda$ value $\lambda_8$. The results are shown with the dashed curves labelled with model-name and MOO (e.g., bmshj2018-hp-8, MOO). As can be seen from Figure \ref{fig:results}, the compression performance improves for all image compression models, especially at lower bitrates, compared to the above discussed results (e.g. bmshj2018-hp-8). Note also that the Mbt2018 model now has much better results as this training adjusts the auto-regressive component of the model to work better for all bitrates.

Next, in the second stage, the QR offsets are introduced into the quantization procedure and the 3-layer NN that generates the QR offsets is trained with a single loss obtained from the sum of the losses with all eight $\lambda$ values. Note that all other weights (obtained from stage 1) of the compression models are frozen in this stage. The results are shown with the curves labelled model-name, MOO and QR-offset (e.g., bmshj2018-hp-8, MOO, QR-offset). As can be seen, the compression performance improves noticeably for Bmshj2018-hp and Mbt2018-mean models compared to the result from stage 1. For the Mbt2018 model, the improvement is insignificant. (Note that if the QR offsets are trained for the models before undergoing MOO training, a similar improvement is obtained, also for Mbt2018 model, as shown by the curves labelled with model-name and QR offsets, e.g., bmshj2018-hp-8, QR-offset.)

In the third stage, the variable rate (VR) quantization is also introduced for the hyper latent $z$ with quantization step-size $\Delta_{z}$ and the 3-layer NN that generates the $\Delta_{z}$ is trained together with all other parameters of the compression models using the MOO training algorithm. The results are shown with the curves labelled model-name, MOO, QR-offset and VR-hp (e.g., bmshj2018-hp-8, MOO, QR-offset, VR-hp). As can be seen, this improves compression performance only slightly for the Bmshj2018-hp and Mbt2018-mean models on top of the results from stage 2. Therefore, VR quantization for hyper latent was not implemented for Mbt2018. 
Note also that while we described a three stage training procedure above (to showcase the individual contributions from the three algorithms), we can confirm that combining all of them into a single MOO-based training stage gives the same results.

% Finally, Figure \ref{fig:results} indicates that the variable-rate compression algorithms by Tong et al. \cite{tong2023qvrf} (and Cui et al. \cite{cui2021asymmetric} as shown in \cite{tong2023qvrf}) show significant compression loss at high bitrates with respect to training multiple models, while the algorithms presented in this paper follow the compression performance of multiple models much closer.

%A summary of computational complexity is as follows. 
The introduced 3-layer fully-connected networks to compute QR offsets and quantization step-sizes $\Delta_z$ %%for the hyper latents 
require 216 and 160 parameters, respectively. The encoding and decoding times increase slightly by a few percents, depending on model and bitrate.

% Can all stages be compiled into a single stage ???

% -----------------------------------------------------------------------------
\section{Conclusions} \label{sec:conc}
% Three algorithms were presented to obtain successful variable bitrate compression performance from a single learned image compression model.The introduced variable bitrate compression algorithms provide compression performance that are close to the compression performance obtainable with multiple models, each trained for a particular bitrate. For hyperprior based models without spatial auto-regressive parts in their entropy models, the compression performance loss is negligible at high bitrates and small at low bitrates. For models with auto-regressive parts in their entropy models, the loss is slightly larger. 

Three algorithms were presented to obtain good variable bitrate compression performance from a single learned image compression model.
The presented results indicate that the combination of the algorithms allows for variable-rate compression performance that follows the compression performance of multiple models closely, in particular at high bitrates, where the compression performance of algorithms by Tong et al. \cite{tong2023qvrf} (and Cui et al. \cite{cui2021asymmetric} as shown in \cite{tong2023qvrf}) show significant compression loss.

\section{References}
\bibliographystyle{IEEEbib}
\bibliography{refs}

\end{document}